\documentclass[reprint,nofootinbib,aps]{revtex4-1}
\usepackage{amsmath,amssymb,amsbsy,amstext,amsthm,simplewick,amsfonts} 
\usepackage{physics}
\usepackage{graphicx} 
\usepackage{dcolumn} 
\usepackage{bm} 
\usepackage{comment}
\usepackage{enumitem}
\usepackage{hyperref}
\hypersetup{
	colorlinks = true,
	linkcolor = blue,
	citecolor = blue,
	urlcolor = blue
}

\def\beq{\begin{equation}}
\def\eeq{\end{equation}}

\begin{document}

\title{Neutrino Effects on Atomic Measurements of the Weinberg Angle}
\author{Mitrajyoti Ghosh$^{1}$}
\email{mghosh2@fsu.edu}

\author{Yuval Grossman$^{2}$}
\email{yg73@cornell.edu}

\author{Chinhsan Sieng$^{2}$}
\email{cs2284@cornell.edu}

\author{Bingrong Yu$^{2}$}
\email{bingrong.yu@cornell.edu}

\affiliation{
	$^1$ Department of Physics, Florida State University, Tallahassee, FL 32306-4350, USA\\
	$^2$ Department of Physics, LEPP, Cornell University, Ithaca, NY 14853, USA
}

\begin{abstract}
We derive a complete expression for the neutrino-mediated quantum force beyond the four-Fermi approximation within the Standard Model. Using this new result, we study the effect of atomic parity violation caused by neutrinos. We find that the neutrino effect is sizable compared to the current experimental sensitivity and can also significantly affect the value of the Weinberg angle measured in atomic systems. This offers a promising method for detecting the neutrino force in the future and facilitates the application of precision atomic physics as a probe for neutrino physics and the electroweak sector of the Standard Model.
\end{abstract}

\maketitle

\section*{Introduction}
Neutrino is the lightest fermion in the Standard Model (SM). As a result, the exchange of a pair of neutrinos can mediate a long-range interaction. This intriguing idea of neutrino-mediated forces can be traced back to Feynman in the 1960s~\cite{Feynman:1996kb}. 
Unlike the usual classical force that is mediated by a boson at the tree level, the neutrino force is a pure quantum effect. It scales as $V_{2\nu}(r)\sim G_F^2/r^5$ in the four-Fermi effective theory~\cite{Feinberg,Feinberg:1989ps,Hsu:1992tg}, where $G_F$ is the Fermi constant. When the nonzero neutrino mass $m_\nu$ is included, $V_{2\nu}$ is sensitive to the nature of the neutrino mass~\cite{Grifols:1996fk}. In particular, the difference caused by the Dirac mass and the Majorana mass is most significant at a distance $r\sim 1/m_\nu$~\cite{Segarra:2020rah,Costantino:2020bei}.
Although the existence of this unique quantum force is a solid prediction of the SM and quantum field theory, it has never been verified experimentally because it is very weak at both long and short distances (see \cite{Horowitz:1993kw,Dzuba:2017cas,Ghosh:2019dmi,Bolton:2020xsm,Munro-Laylim:2022fsv,Ghosh:2022nzo,VanTilburg:2024xib,Ghosh:2024qai} for previous efforts). 

A distinguishing feature of the neutrino force is that it violates parity, which is an exact symmetry of the electromagnetic interaction. Hence, the neutrino force can contribute to parity-violating effects between particles, in particular, in atomic systems~\cite{Ghosh:2019dmi}. Atomic parity violation (APV) provides an important test of the SM at low-energy scales, and the APV observables are also sensitive to fundamental parameters such as the Weinberg angle~\cite{Bouchiat:1997mj}. The APV effect caused by neutrinos has long been believed to be negligible due to the suppression of ${\cal O}(G_F^2)$~\cite{Marciano:1978ed,Marciano:1982mm,Marciano:1983ss}. The purpose of this Letter is to show that this is not the case. 
Unlike the interaction mediated by weak gauge bosons, the neutrino force exhibits a nontrivial radial dependence across the atomic length scale. 
After correctly taking into account the full distribution of the neutrino force in atoms, we find that the APV effect caused by two-neutrino exchange is comparable to the current experimental sensitivity. More importantly, this effect leads to a substantial shift of the Weinberg angle measured from APV experiments.

Technically, one needs to calculate the parity-violating matrix elements induced by $V_{2\nu}$ between atomic states. Using the four-Fermi effective theory, Ref.~\cite{Ghosh:2019dmi} calculated the relevant matrix elements for states with orbital angular momentum $\ell \geq 2$ in hydrogen, which turned out to be too small. 
However, the leading effect in atomic systems (in states with $\ell=0$) caused by neutrinos is sensitive to the short-range behavior of $V_{2\nu}$, a region where the four-Fermi approximation does not apply. 
The short-range behavior of the neutrino force was partially studied in \cite{Xu:2021daf,Dzuba:2022vrv}. However, a complete expression for the SM neutrino force that includes all relevant ultraviolet (UV) contributions and is valid at arbitrary distances is still lacking. This is necessary for quantifying the neutrino contribution to APV and is derived in the present work.

In this Letter, we first obtain a new expression for the neutrino force that is applicable at arbitrary scales. Then we apply it to the study of APV in different atomic systems. As a direct consequence, we calculate the shift of the Weinberg angle measured in these systems caused by neutrinos.
Our result demonstrates that the previously neglected effect from two-neutrino exchange is indeed important for a precision test of the SM at atomic scales. 

This Letter highlights the central ideas and key findings, while additional details are provided in the companion paper \cite{Ghosh:2024ctv}.

\section*{Neutrino forces beyond the four-Fermi theory}
%%%%%%%%%%% % Fig.1  %%%%%%%%%%%%%
\begin{figure*}
	\centering
		\includegraphics[scale = 0.8]{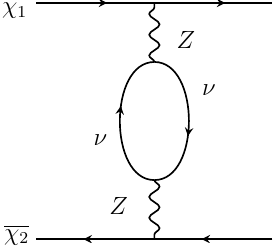}\qquad\qquad
		\includegraphics[scale = 0.8]{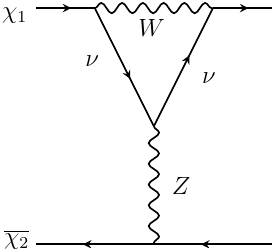}\qquad\qquad
		\includegraphics[scale = 0.8]{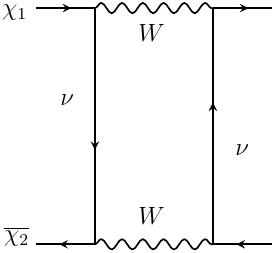}
	\caption{\label{fig:diagrams} Feynman diagrams that contribute to the SM neutrino forces beyond the four-Fermi approximation:  self-energy (left), penguin (middle), and box (right), where $\chi_1$ and $\chi_2$ are arbitrary SM fermions. The coupling to neutrino and $W$ boson exists only when $\chi_i$ is a lepton. If both $\chi_1$ and $\chi_2$ are leptons, there is an additional flipped penguin diagram with the vertex correction at $\overline{\chi_2}$. Note that the box diagram only exists for $\chi_1 = \chi_2$.} 
\end{figure*}
%%%%%%%%%%%%%%%%%%%%%%%%%%%%%%
The four-Fermi approximation breaks down when $r\lesssim \sqrt{G_F}$, a region where the heavy gauge bosons cannot be integrated out. In this case, one needs to calculate the neutrino force in the renormalizable electroweak theory. There are three types of UV diagrams that can contribute to the neutrino force in the SM (see Fig.~\ref{fig:diagrams}): self-energy (SE), penguin (PG), and box diagrams.  In the long-range limit $r\gg \sqrt{G_F}$, all of them yield the known $1/r^5$ form, while at short distances, they lead to a nontrivial radial dependence of the potential that cannot be captured by the four-Fermi effective theory. 

As shown in \cite{Ghosh:2024ctv}, the neutrino force between two SM fermions $\chi_1$ and $\chi_2$ has a generic factorizable form:
\begin{align}
V_{2\nu} (r) \sim G_{V,A}^{\chi_1} \times G_{V,A}^{\chi_2} \times V_0(r)\;,
\end{align}
where $G_{V,A}^{\chi_i}$ are determined by the vector ($V$) or  axial-vector ($A$) couplings between $\chi_i$ and the gauge bosons, while $V_0$ is independent of the couplings of $\chi_i$. Once $V_0$ is obtained, one can easily get the parity-conserving and parity-violating parts of the neutrino force between $\chi_1$ and $\chi_2$ by substituting the corresponding couplings. 

The quantity $V_0$ can be extracted from the amplitudes in Fig.~\ref{fig:diagrams}, which are calculated using quantum field theories. Since $V_0$ can be expressed in terms of the imaginary part of the amplitude~\cite{Feinberg:1989ps}, $V_0$ must be finite according to the optical theorem.  We compute $V_0$ using the dispersion formalism developed in \cite{Feinberg:1989ps}:
\begin{align}
V_0^j(r) = -\frac{1}{4\pi^2 r} \int_0^\infty {\rm Im}\left[{\cal M}^j_{\rm NR}(t)\right] e^{-\sqrt{t}r}\,{\rm d}t\;, 
\end{align}
where ${\cal M}^j_{\rm NR}$  (\text{for $j$ = SE, PG, box}) denotes the Feynman amplitude in the non-relativistic limit. The final results turn out to be (details can be found in \cite{Ghosh:2024ctv}):
\begin{widetext}
\begin{align}
V_0^{\rm SE}(r) &= \left(\frac{g}{4 c_W}\right)^4 \frac{1}{48 \pi^3 r} \left[e^x\left(2+x\right) {\rm Ei}\left(-x\right)+e^{-x}\left(2-x\right) {\rm Ei}\left(x\right)+2\right],\label{eq:VSE}\\
V_0^{\rm PG}(r) &= -\frac{g^4}{2048 \pi^3  c_W^2 r} \int_0^\infty {\rm d}t\,e^{-\sqrt{t}\,r}\,\frac{t\left(2m_W^2 + 3t\right) - 2 \left(m_W^2 + t\right)^2 \log \left(1+t/{m_W^2}\right)} {t^2\left(t-m_Z^2+i\epsilon\right)} \;,\label{eq:VPG}\\
    V_0^{\rm box}(r) &= -\frac{g^4}{512\pi^3 r} \int_0^\infty {\rm d}t\,e^{-\sqrt{t}\,r} \frac{m_W^2 t\left(t+2 m_W^2\right) + \left(t^2-2m_W^4\right)\left(t+m_W^2\right)\log\left(1+t/m_W^2\right)}{t^2 \left(t+2m_W^2\right)^2}\;.\label{eq:Vbox}
\end{align}
\end{widetext}

Here, $g$ is the gauge coupling of the ${\rm SU}(2)_{\rm L}$  group, $c_W\equiv \cos\theta_W$, $s_W\equiv \sin\theta_W $ with $\theta_W$ the Weinberg angle,  $m_{W (Z)}$ is the mass of $W (Z)$ boson, $x\equiv m_Z r$, and ${\rm Ei} (x) \equiv -\int_{-x}^\infty {\rm d}t \  t^{-1} e^{-t}$. The integrals in Eqs.~(\ref{eq:VPG}) and (\ref{eq:Vbox}) can be calculated numerically. Note that in Eq.~(\ref{eq:VPG}), the denominator should be shifted by an infinitesimal imaginary term: $t-m_Z^2 \to t-m_Z^2 + i\epsilon$ with $\epsilon\to 0^+$, to regularize the singularity at the $Z$ boson pole. The real part of the resulting integral is then taken (see Appendix C of Ref.~\cite{Ghosh:2024ctv} for details).

%%%%%%%%%%% % Fig.2  %%%%%%%%%%%%%
\begin{figure*}
	\centering
	\includegraphics[scale = 0.3]{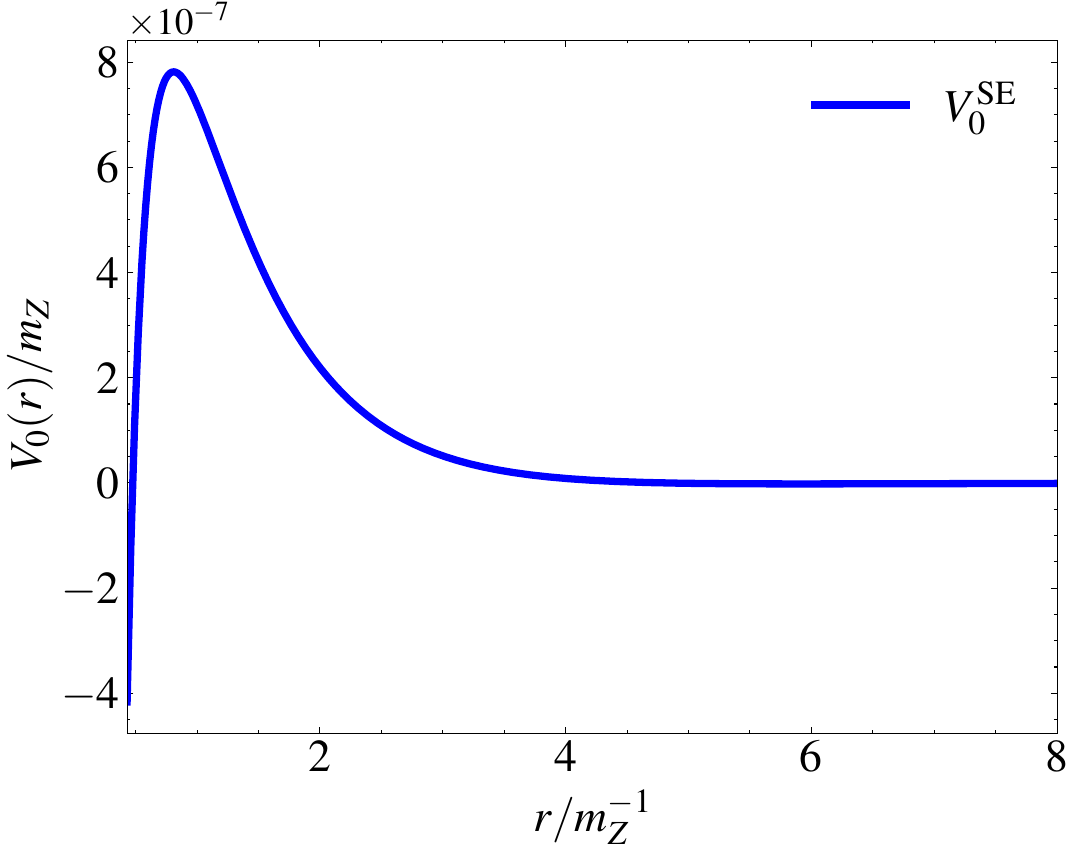}\qquad
	\includegraphics[scale = 0.3]{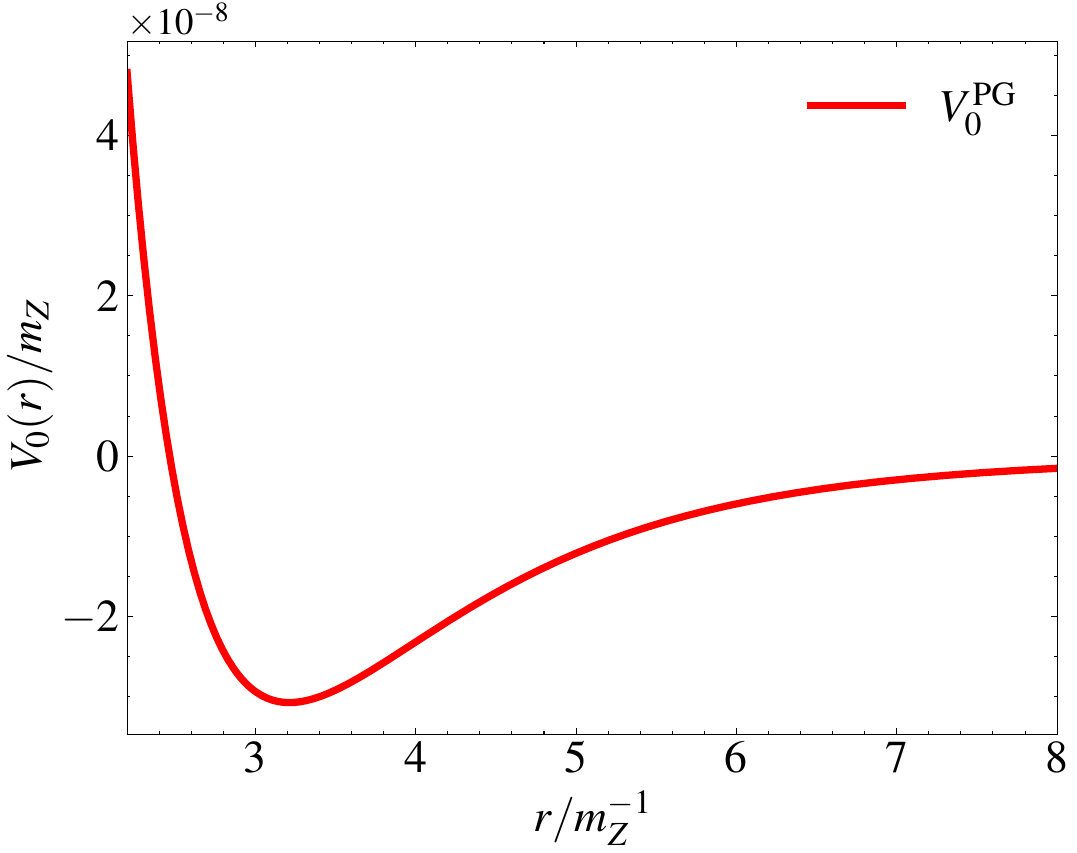}\qquad
	\includegraphics[scale = 0.3]{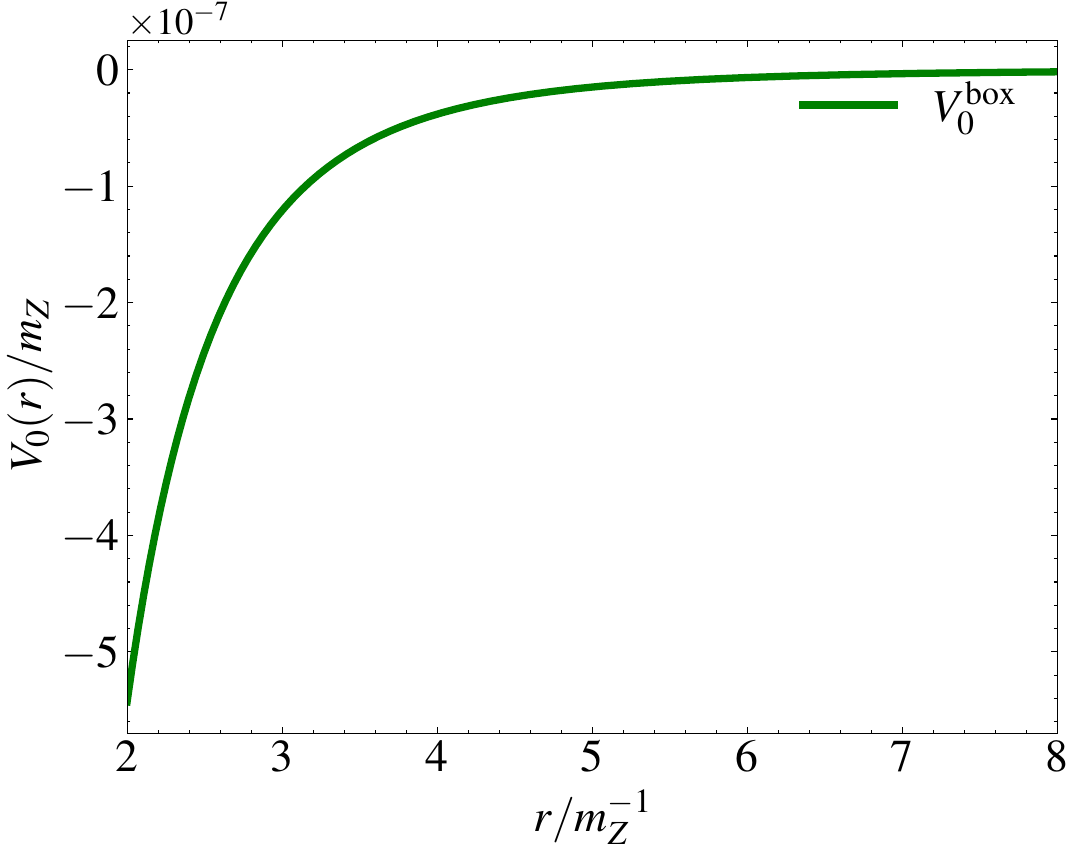}
	\caption{\label{fig:V0} The evolution of the SM neutrino force with distances in the middle range $10^{-1}\lesssim m_Z r \lesssim 10$ according to Eqs.~(\ref{eq:VSE})-(\ref{eq:Vbox}).}
\end{figure*}
%%%%%%%%%%%%%%%%%%%%%%%%%%%%%%

The results in (\ref{eq:VSE})-(\ref{eq:Vbox}) are exact and valid for arbitrary $r$. At long distances $r\gg 1/m_Z$, they reduce to:
\begin{align}
V_0^{\rm SE}(r) \approx \frac{1}{2}V_0^{\rm PG}(r) \approx \frac{1}{4}V_0^{\rm box}(r)\approx -\frac{G_F^2}{16\pi^3 r^5}\;.
\end{align}
In the short-range limit $r\ll 1/m_Z$, we obtain:
\begin{align}
V_0^{\rm SE} (r) &\approx   -\frac{g^4}{3072\pi^3 c_W^4}\frac{\log\left(1/m_Z r\right)}{r}\;,\\
V_0^{\rm PG}(r) &\approx \frac{g^4}{512\pi^3 c_W^2}\frac{\log^2\left(m_W r\right)}{r}\;,\\
V_0^{\rm box}(r) &\approx -\frac{g^4}{256\pi^3 }\frac{\log^2\left(m_W r\right)}{r}\;.
\end{align}

Therefore, from the general expression of the neutrino force in (\ref{eq:VSE})-(\ref{eq:Vbox}),
we recover the known $1/r^5$ form in the long-range limit as expected, while in the short-range limit, the neutrino force scales as $1/r$. Moreover, contributions from penguin and box diagrams are more significant at short distances compared to the self-energy diagram because they are enhanced by $\log^2(m_W r)$.

The behaviors of $V_0^j$ between the long-range and short-range regimes are shown in Fig.~\ref{fig:V0}. We find that $V_0^{\rm SE}$ and $V_0^{\rm PG}$ change signs twice and once, respectively, around $r\sim 1/m_Z$, while $V_0^{\rm box}$ maintains the same sign throughout the range. These nontrivial behaviors are crucial for determining the exact contribution of neutrinos to the APV observables.

\section*{Atomic parity violation from neutrino forces}
Parity is a good symmetry in QED, hence the eigenstates of atomic QED Hamiltonian are states of definite parity. When there exists some parity-violating potential $V_{\rm PV}^j$, these eigenstates are perturbed, and states with different parities can get mixed:
\begin{align}
|A\rangle \to |A\rangle + C_{AB}^j |B\rangle\;,\qquad C_{AB}^j = \frac{\langle B | V_{\rm PV}^j |A\rangle}{E_A-E_B}\;,\label{eq:CAB}
\end{align}
where $|A\rangle$ and $|B\rangle$ are two unperturbed eigenstates with opposite parities and energy eigenvalues $E_A$ and $E_B$. The coefficient $C_{AB}^j$ directly contributes to parity-violating observables such as optical activity in atomic samples; by measuring the optical rotation angle of incident light passing through the sample, one can extract the parity-violating effect caused by $V_{\rm PV}^j$~\cite{Ghosh:2019dmi,Bouchiat:1997mj}.

The leading contribution to parity violation in atoms comes from the tree-level $Z$ exchange between the electron and the nuclei. This effect was observed in Cesium in 1997 with an uncertainty of 0.35\%~\cite{Wood:1997zq}. The neutrino contribution, as a quantum correction to the $Z$ effect, has been ignored for a long time~\cite{Marciano:1978ed,Marciano:1982mm,Marciano:1983ss}. 
%This is because the usual way to extract the APV effect is to calculate the parity-violating amplitude and then take the limit of zero-momentum transfer $q^2 \to 0$ (or equivalently, $m_Z^2\to \infty$) at the amplitude level. In this limit, the ${\cal O}(G_F^2)$ loop amplitudes in Fig.~\ref{fig:diagrams} mediated by two neutrinos are negligible compared to the ${\cal O}(G_F)$ tree-level amplitude mediated by $Z$ and other ${\cal O}(\alpha G_F)$ one-loop amplitudes mediated by multiple gauge bosons, where $\alpha$ is the fine-structure constant. 
%However, the neutrino force actually has a much longer range than the force mediated by heavy gauge bosons. Consequently, its parity-violating effect cannot be correctly captured by ignoring the radial dependence of atomic wavefunctions. The right method is to consider the neutrino force across the entire atomic length scale and convolve it with the atomic wavefunctions, as will be demonstrated below.
In this work, to quantify the neutrino effect in atomic systems,
we carefully take into account the radial dependence of the neutrino force across the entire atomic length scale and convolve it with the atomic wavefunctions, as will be demonstrated below.

We are interested in the APV effect caused by neutrinos in comparison to the $Z$ exchange. This is captured by the following observable:
\begin{align}
\eta_\nu \equiv C_{AB}^\nu / C_{AB}^Z\;,
\end{align}
where $C_{AB}^\nu$ and $C_{AB}^Z$ are the coefficients defined in Eq.~(\ref{eq:CAB}), calculated from the parity-violating matrix elements induced by the neutrino force and the $Z$ force, respectively. The spin prefactors for neutrino forces and that for the $Z$ force are canceled in the ratio $\eta_\nu$.

As an illustration, we take the simple atomic systems of muonium ($\mu^+ e^-$) and positronium ($e^+ e^-$), where the correction from atomic many-body effects can be avoided. The parity-violating neutrino forces in these two systems are found to be:
\begin{widetext}
\begin{align}
    V_\text{$2\nu$,PV}^{e \overline{\mu}}(r) &= \left[\frac{\left(\bm{\sigma}_{\overline{\mu}}-\bm{\sigma}_e\right) \cdot \mathbf{p}_e}{m_e} +\frac{(\bm{\sigma}_{\overline{\mu}} \times \bm{\sigma}_e) \cdot \mathbf{\nabla}}{2m_e}  \right] \left[3 \left(1-4s_W^2\right)  V_0^{\rm SE}(r) - \left(2- 4s_W^2\right) V_0^{\rm PG}(r)  \right]+{\rm h.c.},\\
     V_\text{$2\nu$,PV}^{e \overline{e}}(r) &= \frac{(\bm{\sigma}_{\overline{e}} \times \bm{\sigma}_e) \cdot \mathbf{\nabla}}{m_e}  \left[3 \left(1-4s_W^2\right)  V_0^{\rm SE}(r) - \left(2- 4s_W^2\right) V_0^{\rm PG}(r) + V_0^{\rm box}(r) \right]+{\rm h.c.},
\end{align}
\end{widetext}
where $V_0^j$ are given by Eqs.~(\ref{eq:VSE})-(\ref{eq:Vbox}), $ \mathbf{p}_e$ and $m_e$ are the momentum operator and mass of the electron, $\bm{\sigma}_{\chi}$ denotes the spin operator of $\chi$.  
In addition, let $|A\rangle$ and $|B \rangle$ have orbital angular momenta $\ell$ and $\ell + 1$, respectively. Calculating the matrix elements, we arrive at, for $\ell = 0$ (i.e., the $s$-state):
\begin{align}
	\eta_\nu^{\rm SE} &\approx  -\frac{g^2}{32\pi^2 c_W^2}\approx -0.2\%\;,\\
	\eta_\nu^{\rm PG} &\approx \frac{g^2\left(1-2s_W^2\right)\left(4\pi^2-33\right)}{96\pi^2\left(1-4s_W^2\right)}\approx 4\%\;,\\
	\eta_\nu^{\rm box} &\approx \frac{g^2\left(4+3\pi^2\right)}{256\pi^2\left(1-4 s_W^2\right)} \approx 12\%\;,
\end{align}
where to get numerical values we used $g\approx 0.65$ and $s_W^2\approx 0.24$~\cite{ParticleDataGroup:2024cfk}. We find $\eta_\nu$ does not change significantly for different small values of the principal quantum
number, as the effect is mostly canceled in the ratio.
Note that the relative contributions from the penguin and box diagrams are accidentally enhanced by a factor of $(1-4 s_W^2)^{-1}\sim {\cal O}(10)$, which is due to the vector coupling of charged leptons to the $Z$ boson in the tree-level diagram. The box diagram has the largest contribution because $V_0^{\rm box}$ does not change sign throughout, and there is no suppression from any couplings. 

Therefore, the neutrino contributions in muonium and positronium are given by (for $\ell = 0$):
\begin{align}
	\eta_\nu^{e\overline{\mu}} &= \eta_\nu^{\rm SE} + \eta_\nu^{\rm PG} \approx 4\%\;,\\
	\eta_\nu^{e\overline{e}} &= \eta_\nu^{\rm SE} + \eta_\nu^{\rm PG}  + \eta_\nu^{\rm box} \approx 16\%\;. 
\end{align}

The $Z$ force has an extremely short range ($r\lesssim 1/m_Z$) compared to atomic scales. Therefore, for higher-$\ell$ states, the $Z$ force contribution is heavily suppressed as the atomic wavefunctions are not concentrated at the origin, leading to a relative enhancement of $\eta_\nu$:
\begin{align}
\eta_{\nu}^{\ell =1}  &\sim \log\left(\frac{m_Z^2}{\alpha^2 m_e^2}\right) \eta_{\nu}^{\ell =0}\;,\\
\eta_{\nu}^{\ell \geq 2}  &\sim \left(\frac{m_Z^2}{\alpha^2 m_e^2}\right)^{\ell -1} \eta_{\nu}^{\ell =0}\;.
\end{align}
In this case, the effect of neutrinos can become dominant.

In experiments thus far, the APV effect has only been observed in heavy atoms for $s$-states, typically in Cesium~\cite{Wood:1997zq}. The accidental enhancement of $\eta_\nu$ does not exist in heavy atoms because both the tree and loop contributions are proportional to the vector coupling of neutrons to the $Z$ boson. In this case, we find that for $\ell=0$:
\begin{align}
\eta_\nu^\text{Cs} \sim -\frac{g^2}{16\pi^2} \sim -0.3\%\;.\label{eq:heavyatom}
\end{align}
The magnitude in Eq.~(\ref{eq:heavyatom}) is at the order of a typical loop factor and is sizable compared to the current experimental uncertainty in Cesium (0.35\%)~\cite{Wood:1997zq}.

Note that the $0.3\%$ effect estimated above originates solely from the two-neutrino exchange process, corresponding to the absorptive part of the loop amplitudes in Fig.~\ref{fig:diagrams} obtained by cutting across the neutrino loop. Other one-loop effects of comparable size on APV are not included in this discussion. Nevertheless, the neutrino-force contribution computed here is nontrivial: it cannot be absorbed into any tree-level vertex renormalization, is independent of the renormalization scheme, and was overlooked in previous analyses~\cite{Marciano:1978ed,Marciano:1982mm,Marciano:1983ss}.

\section*{Neutrino forces and the Weinberg angle}
APV experiments have been used to measure the Weinberg angle $\theta_W$ at low-energy scales (which is also the only available method at present). It is connected to the weak charge $Q_W$ through (at the tree level)
\begin{align}
{Q}_W = {\cal Z}\left(1-4 s_W^2\right) -{\cal N}\;,
\label{eq:QW}
\end{align}
where ${\cal Z}$ and ${\cal N}$ are the atomic and neutron numbers of the atom of concern. From the optical rotation angle measurements in APV experiments, one can extract $Q_W$ and thus the value of $\theta_W$ using Eq.~(\ref{eq:QW}). This provides a test of the SM at the atomic scale.

Since we have shown that the neutrino force can contribute to APV observables, it will also affect the value of the Weinberg angle measured in APV experiments. Let $\theta_W$ ($\theta_W'$) be the measured value with (without) the neutrino force taken into account. We obtain
\begin{align}
	\delta s_W^2 &\equiv \left(\sin^2 \theta_W ' - \sin^2 \theta_W\right)/ \sin^2 \theta_W  \nonumber\\
	&\approx \frac{\eta_\nu}{{\cal N}+3{\cal Z}}\left[3{\cal N}+\left(4{\cal N}-3{\cal Z}\right)\left(1-4 s_W^2\right)\right],
\end{align}
where $\eta_\nu$ has been calculated above. Numerically, we arrive at (for $s$-state)
\begin{align}
\delta s_W^2 \approx 
\begin{cases} 
	-0.2\% & \text{muonium} \\
	-0.7\% & \text{positronium}\\
	-0.3\% & \text{Cesium} 
\end{cases}\;.
\end{align}

This shows that neglecting the contribution from neutrinos would make the measured value of the Weinberg angle in APV experiments smaller than its true value. Currently, there is a 2$\sigma$ slight tension between the APV measurement and the SM prediction~\cite{ParticleDataGroup:2024cfk}: the value of $\sin^2\theta_W$ measured from the APV is about 1\% smaller than that predicted by the SM (including the running effect), with an experimental uncertainty of $0.8\%$. Interestingly, including the neutrino effect can help alleviate this tension in the right direction. The ongoing and future APV experiments~\cite{ Antypas_2018,PhysRevA.100.012503,Nanos_2024,Gwinner_2022,Leefer:2014tga, PhysRevA.87.042505,Toh2019} may further confirm this tension with better sensitivities and hopefully, pin down the effect of the neutrino force.

\section*{Conclusions} 
As a fundamental property of the weak force, neutrino interactions violate parity, providing an opportunity to distinguish the neutrino effect from the leading QED effect in atomic systems.
In this work, we derive a complete expression for the neutrino-mediated force beyond the four-Fermi approximation that is valid across all length scales. This enables us to correctly calculate the parity-violating effect in atoms caused by neutrinos. 
%Due to its long-range feature, in general, the neutrino effect cannot be treated simply as a contact correction to that of the tree-level $Z$ exchange without considering the atomic wavefunctions. To get the correct contribution, one must take into account the radial dependence of the neutrino force across the entire atomic scale and convolve it with the atomic wavefunctions. 
Our results show that this effect cannot be neglected, as previously thought, and is comparable to the current APV sensitivity.  Moreover, we show that this effect can lead to a deviation in the measured value of the Weinberg angle. In fact, this deviation is close to the current experimental uncertainty, highlighting the significance of the neutrino effect in atomic systems.
Our work offers a promising method for detecting the neutrino force in the future and, more importantly, reveals the crucial role of neutrinos in precision tests of the SM at the atomic scale.

\begin{acknowledgments}
This work is supported in part by the US Department of Energy grant DE-SC0010102 and the NSF grant PHY- 2309456.
\end{acknowledgments}

\bibliography{ref}

\end{document}